\documentclass[prb,aps,twocolumn]{revtex4}

\usepackage{amsmath,amssymb,dsfont,graphicx,psfrag}
\usepackage{epsfig}
\usepackage{graphicx}% Include figure files

\begin{document}
\title{Nernst Effect as a Signature of Quantum Fluctuations in Quasi-1D Superconductors}

\author{Yeshayahu Atzmon$^{1}$ and Efrat Shimshoni$^{1}$ }
\affiliation{$^{1}$Department of Physics, Bar-Ilan University, Ramat-Gan 52900, Israel }

\date{\today}
\begin{abstract}
We study a model for the transverse thermoelectric response due to quantum superconducting fluctuations in a two-leg Josephson ladder, subject to a perpendicular magnetic field $B$ and a transverse temperature gradient. Assuming a weak Josephson coupling on the rungs, the off-diagonal Peltier coefficient ($\alpha_{xy}$) and the Nernst effect are evaluated as functions of $B$ and the temperature $T$. In this regime, the Nernst effect is found to exhibit a prominent peak close to the superconductor--insulator transition (SIT), which becomes progressively enhanced at low $T$. In addition, we derive a relation to diamagnetic response: $\alpha_{xy}= -M/T_0$, where $M$ is the equilibrium magnetization and $T_0$ a plasma energy in the superconducting legs.

\end{abstract}
\pacs{74.40.-n, 74.25.fg, 74.81.Fa, 71.10.Pm, 05.30.Rt}
\maketitle

\section{Introduction}

\label{sec:intro}

In low-dimensional superconducting (SC) systems (ultra-thin films, wires
and Josephson arrays), fluctuations of the SC order parameter
field lead to broadening of the transition to the SC state, and
give rise to anomalous transport properties in the adjacent normal
phase \cite{SCfluc}. While close to or above the critical
temperature $T_c$ thermally excited fluctuations dominate these
conduction anomalies, quantum fluctuations are expected to
dominate at low temperatures $T\ll T_c$, where superconductivity is
weakened due to, e.g., the effect of a magnetic field, disorder or
repulsive Coulomb interactions. Their most dramatic manifestation
is the onset of a superconductor--insulator transition (SIT) when
an external parameter such as magnetic field or thickness is tuned
beyond a critical point \cite{SITrev,SIT1D}.

A striking signature of the fluctuations regime, which attracted
much attention in the recent years, is the anomalous enhancement
of transverse thermoelectric effects in the presence of a
perpendicular magnetic field $B$. In particular, a substantial
Nernst effect measured far above $T_c$, e.g., in the underdoped
regime of high-$T_c$ superconductors \cite{ong1,ong2} and
disordered thin films \cite{behnia,Pourret}. As the Nernst signal (a
voltage developing in response to a temperature gradient in the
perpendicular direction) is typically small in ordinary metals,
its magnification in such systems has been attributed to the dynamics of thermally excited Gaussian SC
fluctuations \cite{UD,USH,MF,SSVG}, or mobile vortices above a
Kosterlitz-Thouless \cite{KT} transition \cite{PRV}. Theoretical
studies were also extended to the quantum critical regime of
SC fluctuations \cite{BGS}.

Conceptually, the above mentioned theoretical models share a
common intuitive idea: in the phase-disordered, vortex liquid
state (which is qualitatively equivalent to a regime dominated by dynamical Gaussian fluctuations), vortex flow generated parallel to a thermal gradient
($\nabla_y T$) naturally induces an electric field ($E_x$) in the
perpendicular direction. Consequently, the
general expression for the Nernst coefficient
\begin{equation}
\nu\equiv\frac{E_x}{(\nabla_y T) B}=\frac{\rho_{xx}\alpha_{xy}-\rho_{xy}\alpha_{yy}}{B}
\label{Nernst_def}
\end{equation}
is overwhelmingly dominated by
the first term in the numerator, dictated by the off-diagonal Peltier coefficient $\alpha_{xy}$: SC fluctuations typically do not contribute to the second term due to particle-hole symmetry. This is in sharp contrast with ordinary metals, where the two terms almost cancel.
Measurement of the Nernst signal is therefore often regarded as a direct probe of $\alpha_{xy}$,
which is an interesting quantity: while being
a transport coefficient, it is intimately related to thermodynamic properties. In particular, it was found (both experimentally and theoretically) to be proportional to the diamagnetic response \cite{ong2,USH,PRV}: $\alpha_{xy}\sim -M/T$. In the clean limit (i.e. for Galilean invariant systems), it was shown to encode the entropy per carrier \cite{CHR,BO,SRM}.

\begin{figure}
\includegraphics[width=0.9\linewidth]{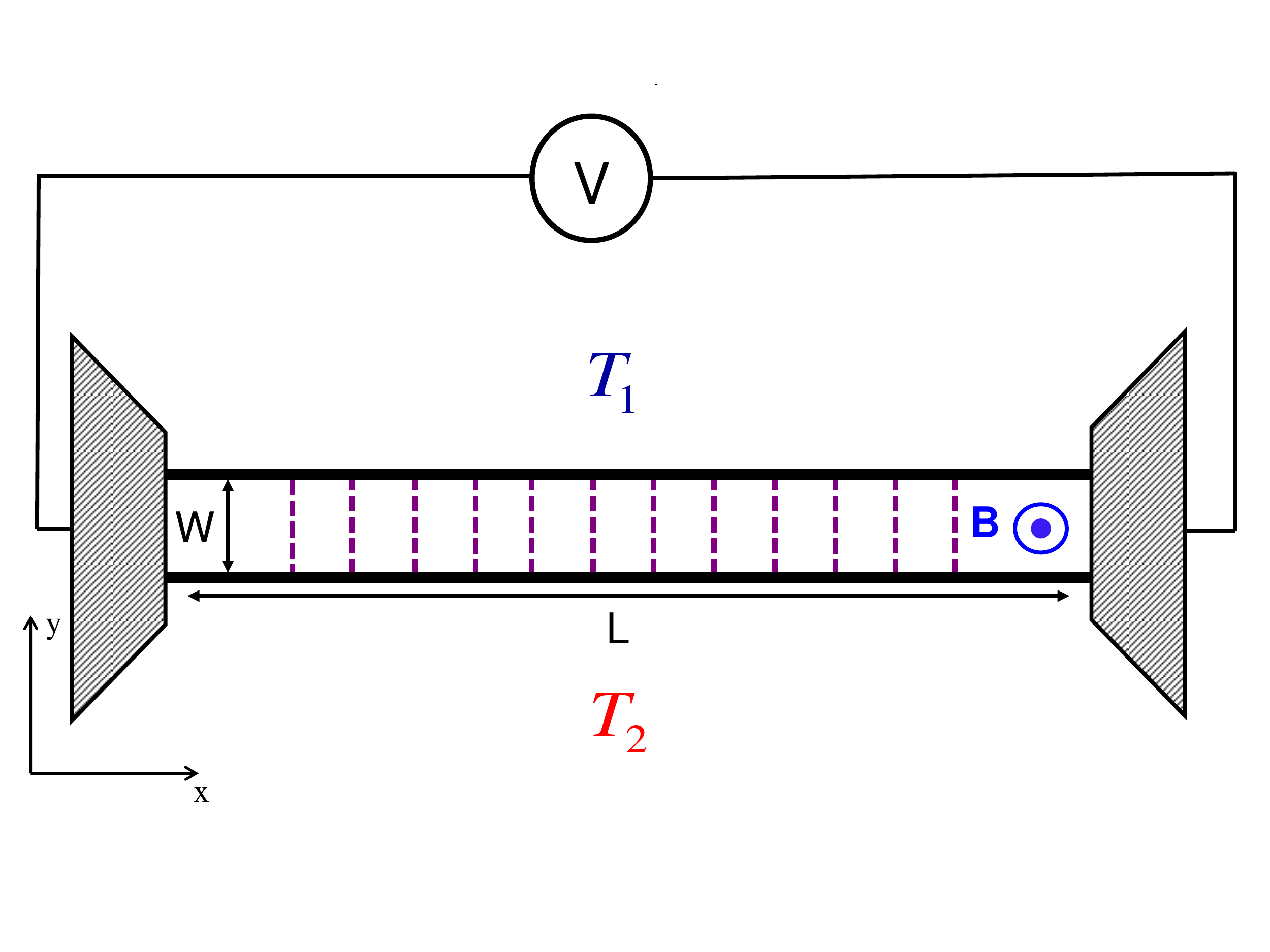}
\caption{(color online)
A scheme for measurement of the Nernst effect in a Josephson ladder subject to a magnetic field $B$ perpendicular to the plane, and a temperature difference between the top ($T_1$) and bottom ($T_2$) SC wires. Dashed lines represent Josephson coupling.
 \label{fig1} }
\end{figure}

Note, however, that even in the case where Eq. (\ref{Nernst_def}) is dominated by the first term,
the overall contribution to $\nu$ is not determined by $\alpha_{xy}$ alone, but rather by its product with the electric resistivity $\rho_{xx}$. Observation of a large Nernst signal therefore necessitates a reasonably resistive normal state. This indicates that a large Nernst signal is a subtle effect: on one hand it requires the presence of superconducting fluctuations, and on the other hand requires the superconducting fluctuations to be {\it dynamic} in order to produce a sizable voltage drop. A conjunction of these competing tendencies occurs in the fluctuations dominated regime. Moreover, the Nernst effect is expected to serve as a sensitive probe of a SIT. It should be pointed out, however, that when the normal state adjacent to the SC transition is an insulator, $\alpha_{xy}$ can {\it not} be directly deduced from $\nu$: unlike most cases studied in the literature thus far, $\rho_{xx}$ can not be assumed constant. Rather, it possesses a significant dependence of its own on the deviation from the critical point, and on $T$ (in particular, a {\it divergence} in the $T\rightarrow 0$ limit). As a result, although $\alpha_{xy}$ is bound to vanish for $T\rightarrow 0$ by the third law of thermodynamics, $\nu$ may in principle approach a finite value in this limit.

Motivated by the above general observations,
in the present paper we develop a theory for the transverse
thermoelectric coefficients and their relation to diamagnetism in
the quasi one-dimensional (1D) superconducting device depicted in
Fig. \ref{fig1}, in which the geometry dictates an appreciable
Nernst effect in the fluctuations-dominated regime. The device
considered is a two-leg Josephson ladder subject to a
perpendicular magnetic field $B$, where a small temperature
difference between the legs induces voltage along the ladder due
to transport of vortices across the junction \cite{glazman}. At low $T$, one
expects vortex transport to be dominated by quantum tunneling.
This system serves as a minimal setup for observing transverse
thermoelectric effects \cite{Jnernst}. The relative simplicity of the model describing
the quantum dynamics of SC fluctuations allows an explicit
evaluation of $\alpha_{xy}$, $\nu$ and the magnetization density
$M$ in a wide range of parameters. In particular, we investigate
their behavior when the wires parameters are tuned through a SIT,
and find a prominent peak in $\nu$ close to the transition, which
becomes progressively enhanced \cite{3rd_law} at low $T$. We further confirm the
proportionality relation between $\alpha_{xy}$ and $-M$, however
the prefactor is $1/T_0$, with $T_0$ the plasma energy scale,
rather than $1/T$ as found in the 2D case \cite{USH,PRV}.

The rest of the paper is organized as follows: in Sec. \ref{sec:model} we set up the model.
In Sec. \ref{sec:thermoelectric} we detail our derivation of the transverse thermoelectric coefficients $\alpha_{xy}$
(subsection \ref{sec:alpha_xy}) and $\nu$ (subsection \ref{sec:nu}). In Sec. \ref{sec:M} we derive
a relation between $\alpha_{xy}$ and the diamagnetic magnetization ($-M$).
Finally, in Sec. \ref{sec:discussion} we summarize our main results and conclusions.

\section{The Model}
\label{sec:model}

We consider the system indicated in Fig. \ref{fig1}, which consists of two SC
wires of length $L$ parallel to the $x$ direction separated by a
thin insulating layer of width $W$, which allows a weak Josephson
coupling $J$ per unit length. In each of the separate wires
($n=1,2$), the 1D quantum dynamics of fluctuations in the phase of
the SC condensate is governed by a Hamiltonian of
the form (in units where $\hbar =1$)
\begin{equation}
H_n = \frac{v}{{2\pi }}\int_{ - \frac{L}{2}}^{\frac{L}{2}} d x\left[ {K{{({\partial _x}{\theta _n})}^2 + }\frac{1}{K}{{({\partial _x}{\phi _n})}^2}} \right]\; ;
  \label{Hn}
\end{equation}
here $\phi _{n}(x)$ is the collective phase field, and $\theta
_{n}(x)$ its conjugate field (obeying $[\phi _{n}(x),\partial
_{x}\theta _{n}(x^{\prime })]=i\pi \delta (x^{\prime }-x)$) which
denotes Cooper pair number fluctuations. This model can be viewed as
describing, e.g., the continuum limit of a Josephson chain
\cite{SIT1D}, where the Josephson coupling ($E_J$) and charging
energy ($E_c$) between adjacent SC grains are related to the
parameters of $H_n$ by $K=\sqrt{E_c/E_J}$ and $v=\sqrt{E_cE_J}\pi
a$, with $a$ the grain size. Note that Eq. \ref{Hn} is a low-energy approximation of the quantum fluctuations. As will be shown later, to get non-trivial transverse thermoelectric effects we will need to keep corrections to $H_n$ involving, e.g., higher derivatives of the fields. Such corrections
take into account, for example, fluctuations in the current density within the finite width of the wires ($\sim a$), and coupling of the collective modes to microscopic degrees of freedom.

The Josephson coupling between the
wires is given by
\begin{equation}
H_J=-J\int_{ - \frac{L}{2}}^{\frac{L}{2}} d x\, \cos \{{\phi_1} -
{\phi_2} - qx\}
  \label{HJ}
\end{equation}
where $q$ is the deviation of the vortex density in the junction
area from a commensurate value:
\begin{equation}
q=2\pi\left(\frac{WB}{\Phi_0}\;{\rm mod}\;\frac{1}{a}\right)
 \label{q_def}
\end{equation}
in which (for $\hbar=c=1$) $\Phi_0=\pi/e$ is the flux quantum.
Assuming the hierarchy of scales $Ja\ll T\ll v/a$ (with $T$ an
average temperature of the system), $H_J$ [Eq. (\ref{HJ})] can be
treated perturbatively. Note that the first inequality justifies
this approximation for an arbitrary value of the Luttinger
parameter $K$ in Eq. (\ref{Hn}): for $K<2$ and sufficiently small
$q$, the Josephson term becomes relevant \cite{Giamarchi}, and induces
a SC phase where fluctuations in the relative phase between the
wires are gapped \cite{OG,AS} in the $T\rightarrow 0$ limit. In
turn, additional perturbations such as inter-wire charging energy
\cite{OG,AS} and disorder \cite{GS} generate a transition to an
insulating $T\rightarrow 0$ phase for sufficiently large $K$.
Since, as shown below, in both extreme phases the Nernst effect is
strongly suppressed, we focus our attention on the {\it intermediate}
$T$ regime, where temperature exceeds the energy scale associated with all these perturbations.

In addition to $H_J$, we introduce a weak
backscattering term due to disorder of the form
\begin{eqnarray}
\label{HD} H_{D}&=&\sum_{n=1,2}\int dx\zeta_n (x)\cos \{2\theta
_{n}(x)\}\; ,\\ \overline{\zeta_n (x)} &=& 0\; ,\quad
\overline{\zeta_n (x)\zeta_{n^{\prime}} (x^{\prime })} =D\delta
(x-x^{\prime })\delta_{n,n^{\prime}} \nonumber
\end{eqnarray}
where overline denotes disorder averaging. This term generates the
leading contribution to the resistivity $\rho_{xx}$, and thus to
$\nu$ via Eq. (\ref{Nernst_def}).

\section{The Transverse Thermoelectric Coefficients}
\label{sec:thermoelectric}

We next consider a temperature difference $\Delta T=T_1-T_2$
between the top and bottom wires (see Fig. \ref{fig1}), each
assumed to be at equilibrium with a separate heat reservoir. As a result of the transverse Peltier effect, a
current is induced along the legs of the ladder. Alternatively, if one maintains open boundary conditions,
a voltage develops along the wires due to the Nernst effect. Below we derive the corresponding coefficients.

\subsection{The Off Diagonal Thermoelectric Coefficient $\alpha_{xy}$}
\label{sec:alpha_xy}

We first evaluate the electric current $I_x$ induced along the ladder for
$\Delta T \ll T \equiv \frac{{{T_1} + {T_2}}}{2}$, which yields the transverse Peltier coefficient
\begin{equation}
\alpha_{xy}=\frac{I_x}{\Delta T}\; .
 \label{alpha_xy_def}
\end{equation}
Alternatively, implementing the Onsager relations \cite{USH}, one can deduce it from the coefficient
$\tilde\alpha_{xy}=T\alpha_{xy}$, which dictates the heat
current $I^{(h)}_x$ generated along the device in response to a voltage difference $V_y$
between the two wires. We show explicitly below that the result of both
calculations is indeed the same.

The electric current is given by the expectation value
\begin{equation}
I_x = 2e\pi \left\langle \dot {\theta }_1 +\dot {\theta
}_2\right\rangle
 \label{Ix_def}
\end{equation}
where the current operators $\dot{ {\theta }}_n
(x,t)=i[H,{\theta_n }]$
($H=H_0+H_J$ where $H_0\equiv H_1+H_2$) are evaluated
perturbatively in $H_J$ [Eq. (\ref{HJ})] using the interaction representation.
The leading contribution to $I_x$ arises from the second order:
\begin{eqnarray}
& &\dot{ {\theta }}_n (x,t)= U(t){{\dot {\theta}_n^{(0)}
}(x,t)}{U^\dag }(t)\quad{\rm where} \nonumber \\ \label{I_n}&
&\dot {\theta}_n^{(0)} (x,t) = \frac{v}{K}{\partial_x}{\phi
_n}(x,t)\; ,\\
& &U(t)=  1 + {\mathop {i\smallint }\limits_{ - \infty }
^t}d{t_1}{H_J}({t_1}) - {\mathop \smallint \limits_{ - \infty }
^t}d{t_1}{\mathop \smallint \limits_{ - \infty }
^{{t_1}}}d{t_2}{H_J}({t_1}){H_J}({t_2})\; .
 \nonumber
\end{eqnarray}
Employing Eq. (\ref{HJ}) and inserting the resulting expressions for $\dot{ {\theta }}_n$ in Eq. (\ref{Ix_def}), we obtain
\begin{widetext}
\begin{equation}
\begin{array}{*{20}{l}}
{{I_x} = \frac{{{\pi evJ^2}}}{2}\int\limits_{ - \infty }^t {d{t_1}\int\limits_{ - \infty }^{t_1} {d{t_2}} } \int\limits_{ - \frac{L}{2}}^{\frac{L}{2}} {d{x_1}\int\limits_{ - \frac{L}{2}}^{\frac{L}{2}} {d{x_2}} } \sin \left[ {q({x_1} - {x_2})} \right]
i\Im m\left[ {{{ {{e^{ - \frac{K}{2}{F_{T_1}}\left( {{x_1} - {x_2};{t_1} - {t_2}} \right)}}} }}{{ {{e^{ - \frac{K}{2}{F_{T_2}}\left( {{x_1} - {x_2};{t_1} - {t_2}} \right)}}} }}} \right]}\\
{\left[ {\left\{ {{{\left. {{\partial _x}{F_{T_1}}(x - {x_1};t - {t_1}) + {\partial _x}{F_{T_1}}({x_2} - x;{t_2} - t)} \right\}}} - \left\{ {{{\left. {{\partial _x}{F_{T_2}}(x - {x_1};t - {t_1}) + {\partial _x}{F_{T_2}}({x_2} - x;{t_2} - t)} \right\}}}} \right.} \right.} \right.}\\
 - \left. {\left\{ {{{\left. {{\partial _x}{F_{T_1}}(x - {x_2};t - {t_1}) + \partial_x{F_{T_1}}({x_1} - x;{t_2} - t)} \right\}}} + } \right.\left\{ {{{\left. {{\partial _x}{F_{T_2}}(x - {x_2};t - {t_1}) + {\partial _x}{F_{T_2}}({x_1} - x;{t_2} - t)} \right\}}}} \right.} \right]\; ,
\end{array}
\label{Ix_int}
\end{equation}
where we use the Boson correlation function $F_T(x;t)\equiv \frac{1}{K}\langle[\phi_n(x,t)-\phi_n(0,0)]^2\rangle$ at fixed temperature $T$: \cite{Giamarchi}
\begin{equation}
{F_T}(x;t) = \frac{1}{2}\log \left[ {\frac{{{v^2}}}{{{\pi ^2}{a ^2}{T^2}}}\left\{ {\left. {\sinh \left[ {\pi T\left( {\frac{x}{v} - t + i\epsilon\, {\rm sign}(t)} \right)} \right]\sinh \left[ {\pi T\left( {\frac{x}{v} + t - i\epsilon \,{\rm sign}(t)} \right)} \right]} \right\}} \right.} \right]
\label{FT_def}
\end{equation}
\end{widetext}
in which $\epsilon\sim a/v$ is associated with the short-distance cutoff.

Performing the integral first over the center-of-mass coordinate $x_c=\frac{x_1+x_2}{2}$ in Eq. (\ref{Ix_int}),
and taking the limit $\epsilon\rightarrow 0$ in Eq. (\ref{FT_def}), it is easy to see that, since
$\Im m\{F\}=\pi$ is independent of $T$, the resulting expression actually vanishes.
This follows from the Lorentz invariance of the model for phase-fluctuations, Eq. (\ref{Hn}). As we discuss further in Sec. \ref{sec:discussion}, this behavior is in fact quite characteristic: deviations from a linear energy-momentum dispersion are required to get a finite $\alpha_{xy}$. In the present case, a non-vanishing result (of order $\epsilon$) would emerge if $\epsilon$ in Eq. (\ref{FT_def}) is kept finite. This signifies that the leading contribution to $\alpha_{xy}$ arises from physics on scales of the short-distance cutoff, which depends on microscopic details. We therefore need to include such corrections to Eq. (\ref{Hn}), namely terms which violate Lorentz invariance: as will be elaborated in Sec. \ref{sec:discussion}, such terms are indeed necessary to provide the Josephson vortices in this system with entropy.

Tracing back to the underlying microscopic theory of SC devices, Eq. (\ref{Hn}) is an effective Hamiltonian for the collective fields $\phi_n$, $\theta_n$ arising to leading order in a gradient expansion. A variety of higher energy corrections, allowed by the symmetries of the problem, are present in any physical system. In particular, as a concrete example, corrections to the Josephson Hamiltonian which hybridize phase and charge fluctuations have been derived in earlier literature \cite{ESA,SGF} for a single junction. When incorporated in the continuum limit of a model for Josephson array, these yield higher order derivatives, e.g. a term of the form \cite{SGF}
\begin{equation}
\label{HJcorr}
H_n^{corr}=\frac{\mathcal{C}va}{2\pi}\int_{ - \frac{L}{2}}^{\frac{L}{2}}dx\,[(i{\partial _x}{\phi_n})({\partial_x^2}{\theta_n})+h.c.]
\end{equation}
where $\mathcal{C}$ is a dimensionless constant. This adds a correction to the current operator in the wire $n$
[Eq. (\ref{I_n})] of the form
\begin{equation}
\label{I_ncorr}
\delta\dot {\theta}_n^{(0)}=i\mathcal{C}va({\partial_x^2}{\theta_n})\approx \frac{i\mathcal{C}a}{K}\partial_t{\partial _x}{\phi_n}
\end{equation}
where in the last approximation we have used the leading term in the equation of motion for $\phi_n$.
Inserting these corrections to $\dot{\theta}_1$, $\dot{\theta}_2$ into Eqs. (\ref{Ix_def}), (\ref{I_n}) we obtain
\begin{widetext}
\begin{equation}
\begin{array}{*{20}{l}}
{{I_x} \approx -\frac{{{\pi e\mathcal{C}aJ^2}}}{2}\int\limits_{ 0 }^{\infty} {d{t}} } \int\limits_{ - \frac{L}{2}}^{\frac{L}{2}} {d{x_c}\int\limits_{ - L}^{L} {d{x}} } \sin \left[ {qx} \right]
\Im m\left[ {{{ {{e^{ - \frac{K}{2}{F_{T_1}}\left( {x;t} \right)}}} }}{{ {{e^{ - \frac{K}{2}{F_{T_2}}\left( {x;t} \right)}}} }}} \right]\\
{\left[ {\left\{ {{{\left. {{\partial _x}{F_{T_1}}\left(-x_c - \frac{x}{2};0\right) + {\partial _x}{F_{T_1}}\left(x_c - \frac{x}{2};-t\right)
} \right\}}} - \left\{ {{{\left. {{\partial _x}{F_{T_2}}\left(-x_c - \frac{x}{2};0\right)
 + {\partial _x}{F_{T_2}}\left(x_c - \frac{x}{2};-t\right)
} \right\}}}} \right.} \right.} \right.}\\
 - \left. {\left\{ {{{\left. {{\partial _x}{F_{T_1}}\left(-x_c + \frac{x}{2};0\right)
  + \partial_x{F_{T_1}}\left(x_c + \frac{x}{2};-t\right)
 } \right\}}} + } \right.\left\{ {{{\left. {{\partial _x}{F_{T_2}}\left(-x_c + \frac{x}{2};0\right)
+ {\partial _x}{F_{T_2}}\left(x_c + \frac{x}{2};-t\right)
} \right\}}}} \right.} \right]\; ,
\end{array}\nonumber
\end{equation}
\begin{eqnarray}
\label{Ix_int_corr}
&\approx &-(\Delta T)2\pi^2e\mathcal{C}aJ^2\int\limits_{ 0 }^{\infty} {d{t}}\int\limits_{-\infty}^{\infty} {d{x}}\,x\sin \left[ {qx} \right]
\Im m \left\{\chi(x,t)\right\}\; ,\\
& & \chi(x,t)=\frac{(\pi aT/v)^K}{\left(\sinh \left[ {\pi T\left( {\frac{x}{v} - t + i\epsilon} \right)} \right]\sinh \left[ {\pi T\left( {\frac{x}{v} + t - i\epsilon } \right)} \right] \right)^{K/2}} \nonumber
\end{eqnarray}
\end{widetext}
where in the last step we have assumed further $\Delta T\ll T$ and $L\rightarrow\infty$, and inserted the explicit expression for ${F_T}$ Eq. (\ref{FT_def}). Employing the definition Eq. (\ref{alpha_xy_def}),
we evaluate the remaining integrals and finally get
\begin{widetext}
\begin{equation}
%\begin{array}{lll}
\label{alpha_xy_general}
\alpha_{xy} = - \frac{e{(\pi J)^2}a^3\mathcal{C}}{4v^2}
\sin \left( {\frac{{\pi K}}{2}} \right){\left( {\frac{{2\pi a T}}{v}} \right)^{K-2}}{\partial _q}\left\{
\left|B\left( {i\frac{{vq}}{{4\pi T}} + \frac{K}{4},1 - \frac{K}{2}} \right)\right|^2\right\}
%{\partial _q}\left\{ {\left. {B\left( { - i\frac{{vq}}{{4\pi T}} + \frac{K}{4},1 - \frac{K}{2}} \right)
%B\left( {i\frac{{vq}}{{4\pi T}} + \frac{K}{4},1 - \frac{K}{2}} \right)} \right\}} \right.
%\end{array}
\end{equation}
\end{widetext}
where $B(z,w)$ is the Beta function \cite{GRbook}. Note that the resulting $\alpha_{xy}(T)$ exhibits a power-law $T$-dependence, which indicates an apparent divergence in the $T\rightarrow 0$ for sufficiently small $K$. Since $\alpha_{xy}$ is known to be proportional to the entropy of carriers, such behavior would violate the third law of thermodynamics. We emphasize, however, that this is an artefact of the approximation leading to Eq. (\ref{alpha_xy_general}), which assume a finite temperature and in particular $J\ll T$. While the true $T\rightarrow 0$ is beyond the scope of the present theory, we speculate that due to an opening of a gap (either superconducting or insulating, depending on the value of $K$), the coefficient $\alpha_{xy}$ is suppressed as expected.

We now consider the alternative setup where an electric voltage $V_y$ is imposed between the top and bottom wires (at uniform $T$), and evaluate the heat current $I^{(h)}_x$ induced along the ladder. For $J=0$, but accounting for the corrections $H_n^{corr}$ [Eq. (\ref{HJcorr})], the local heat current operator is given by \cite{RA}
\begin{equation}
J_h^{(0)}=v^2\sum_{n=1,2}\partial_x\theta_n
\left({\partial_x}{\phi_n}+i\mathcal{C}\frac{a}{v}\partial_t{\partial_x}{\phi_n}\right)\; .
\label{Jh_def}
\end{equation}
The voltage bias corresponds to a difference in chemical potentials in the two legs, $\mu_{1,2}=\pm eV_y$, which introduce constant shifts of $\partial_x\theta_{1,2}$ by $\pm \pi eV_y/vK$.
The heat current $I^{(h)}_x=\left\langle U(t)J_h^{(0)}{U^\dag }(t)\right\rangle$
[with $U(t)$ expanded to second order in $H_J$ as in Eq. (\ref{I_n})], is hence given by
\begin{equation}
I^{(h)}_x=e\pi V_y\left\langle U(t)(\dot {\theta}_1^{(0)}-\dot {\theta}_2^{(0)}){U^\dag }(t)\right\rangle\; .
\label{Ih}
\end{equation}
The resulting expression coincides with $(V_yT/\Delta T)I_x$. We thus confirm that $\tilde\alpha_{xy}=I^{(h)}_x/V_y=T\alpha_{xy}$, as required by Onsager's relation.

\subsection{The Nernst Coefficient $\nu$}
\label{sec:nu}

To derive the Nernst coefficient, we next employ Eq. (\ref{Nernst_def}) noting that within our level of approximations, $\alpha_{yy}$ and $\rho_{xy}$ vanish due to particle-hole symmetry. The Nernst signal in the setup depicted in Fig. \ref{fig1}, defined as $\nu=|V/\Delta TB|$, is hence determined by the product of $\alpha_{xy}$ [Eq. (\ref{alpha_xy_general})] and the longitudinal resistance of the ladder $R_{xx}$. To leading order in $H_D$ [Eq. (\ref{HD})] \cite{Giamarchi,SIT1D,GS},
\begin{equation}
{R_{xx}} = \frac{{\pi^3 LDa^2}}{{2{e^2}v^2}}\cos \left( {\frac{\pi }{K}} \right)B\left( {\frac{1}{K},1 - \frac{2}{K}} \right){\left( {\frac{{2\pi a T}}{v}} \right)^{\frac{2}{K}-2}}\; .
\label{rho_xx}
\end{equation}
At low magnetic field such that $q=2\pi WB/\Phi_0\ll T/v$, this yields an expression for $\nu\approx\alpha_{xy}R_{xx}/B$ of the form
\begin{equation}
\nu\approx\nu_0{\mathcal F}(K)\left( {\frac{{2\pi a T}}{v}} \right)^{K+\frac{2}{K}-6}
\label{nu}
\end{equation}
where the constant prefactor $\nu_0\propto LDJ^2$ and
\begin{equation}
{\mathcal F}(K)\equiv\frac{\Gamma^2\left(\frac{K}{4}\right)\Gamma\left(1-\frac{K}{2}\right)\Gamma\left(\frac{1}{K}\right)
\left\{\psi^\prime\left(\frac{K}{4}\right)-\psi^\prime\left(1-\frac{K}{4}\right)\right\}}
{2^{2/K}\Gamma^2\left(1-\frac{K}{4}\right)\Gamma\left(\frac{K}{2}\right)\Gamma\left(\frac{1}{K}+\frac{1}{2}\right)}
\label{F(K)}
\end{equation}
[$\Gamma(z)$, $\psi^\prime(z)$ are the Gamma and Trigamma functions, respectively \cite{GRbook}].

\begin{figure}
\includegraphics[width=1.0\linewidth]{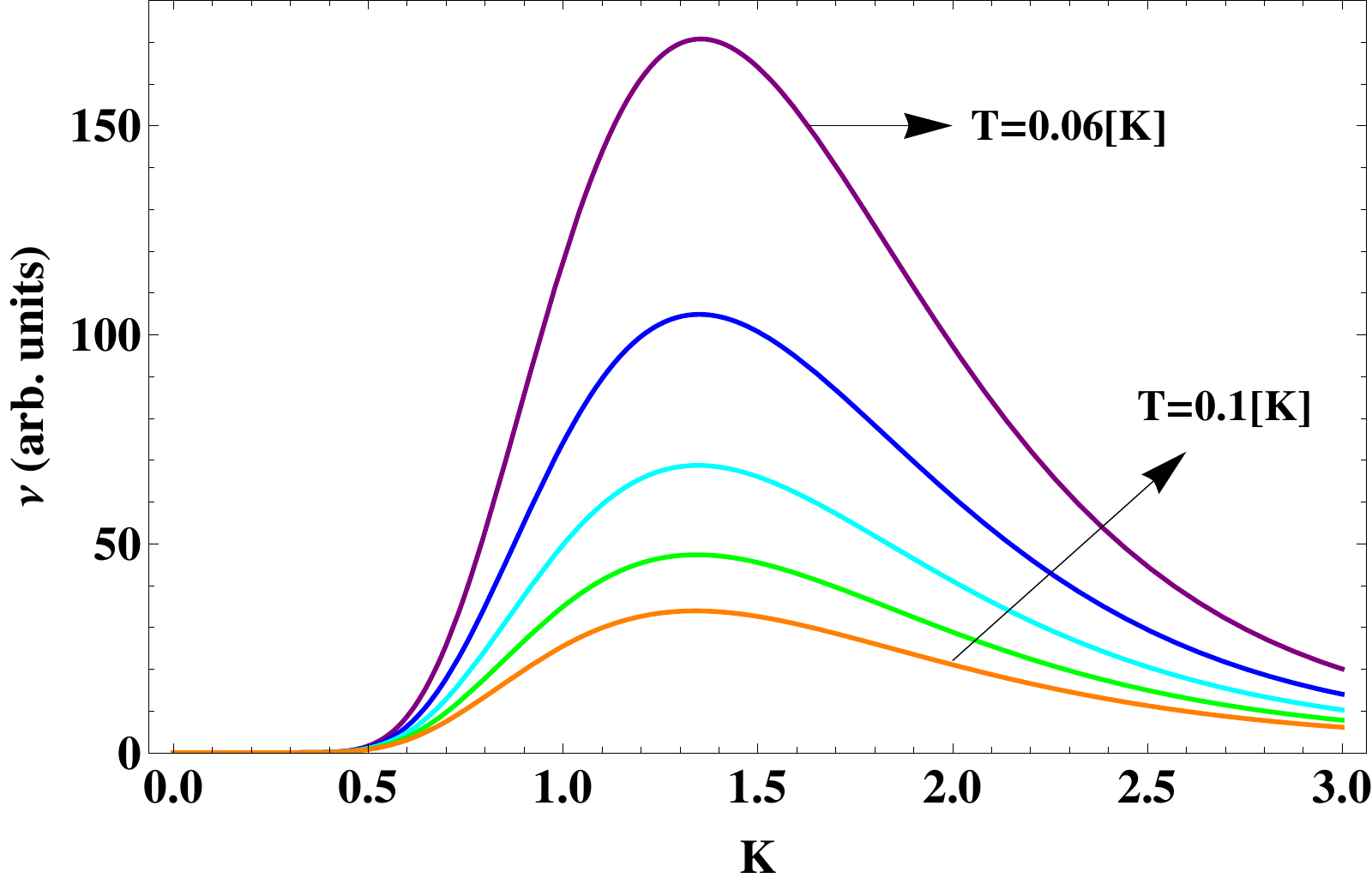}
\caption{(color online)
Isotherms of $\nu$ as a function of $K$ for $v/2\pi a=1$K, $T=0.06,0.07,0.08,0.09,0.1$K.
 \label{fig2} }
\end{figure}

The resulting dependence of $\nu$ on $K$, the parameter which tunes the SIT in the SC wires,
is depicted in Fig. \ref{fig2} for different values of $T\ll v/ a$, and for low magnetic fields where $vq\ll T$.
In this regime, $\nu$ exhibits a pronounced maximum at $K^\ast(T)$, slightly below the transition from SC to
insulator \cite{SIT1D,zaikin}($K_c=2$). As $T$ is lowered, the peak becomes progressively enhanced
and $K^\ast\sim\sqrt{2}$ as dictated by the rightmost exponential factor in Eq. (\ref{nu}).
This non-monotonous behavior can be traced back to the competition between electric resistance
(which signifies the rate of phase-slips), and $\alpha_{xy}$ (which signifies the strength of diamagnetic response).

\section{The Magnetization $M$}
\label{sec:M}

We next investigate the relation of $\alpha_{xy}$ to the diamagnetic magnetization density $M =  - (1/LW)(\partial F  /{\partial B})$, assuming that the $B$-dependence of the free energy $F$ is primarily restricted to the flux in the junction area, i.e. the parameter $q$ in Eq. (\ref{HJ}). To leading order in $H_J$ and at low magnetic field $B=(\Phi_0/2\pi W)q$,
\begin{eqnarray}
\label{FJ}
M &\cong &\frac{2\pi}{\Phi_0}\frac{T}{{2L }}\partial_q{\left\langle {{ {S_J}^2}} \right\rangle _0}\; , \\ \nonumber
S_J &=&- J\int\limits_{ - \frac{L}{2}}^{\frac{L}{2}} dx\int\limits_0^{\frac{1}{T}   } {d\tau }\cos \left\{ {{\phi _1}({x},{\tau}) - {\phi _2}({x},{\tau }) - q{x}} \right\}\; .
\end{eqnarray}
%where $\langle\;\rangle _0$ is evaluated with respect to $H_0=sum_n H_n$.
The resulting expression for $M$ is identical to Eq. (\ref{alpha_xy_general}) up to a constant prefactor. We thus obtain the relation
\begin{equation}
\alpha_{xy}=-\frac{M}{T_0}\; ,\quad T_0\equiv \frac{v}{\pi a\mathcal{C}}\sim\sqrt{E_cE_J}
\label{alpha_xy_M}
\end{equation}
where the last proportionality relation associates the energy scale $T_0$ with the plasma energy in the Josephson chain forming the legs of the ladder. This resembles the linkage pointed out in earlier literature, except the thermal energy scale $T$ in the prefactor (as obtained for full-fledged 2D systems) is replaced here by the characteristic scale of quantum dynamical phase-slips.

\section{Discussion}
\label{sec:discussion}

To summarize, we studied the transverse thermoelectric coefficients due to quantum SC fluctuations in a Josephson two--leg ladder, and their relation to diamagnetism. Most importantly, we predict a large Nernst signal, particularly at moderately low temperatures ($Ja\ll T\ll T_0$) where a pronounced peak is predicted close to the SIT. This behavior reflects a subtle interplay between diamagnetism (favored in the SC phase), and dynamical phase-fluctuations (which proliferate in the insulator).

A crucial step in the derivation of the leading contribution to the transverse Peltier coefficient $\alpha_{xy}$ relies on a correction to the Hamiltonian which violates Lorentz invariance of the model for quantum phase fluctuations in the SC wires. Such terms are indeed necessary to couple the charge current to the thermal current: the correlation between them is at the heart of the Peltier effect.
This point can be understood also from a different angle, implementing the language of vortex physics: it is in fact possible to view $\alpha_{xy}$ as a dual of the ordinary (longitudinal) thermopower $Q=\Delta V/\Delta T$, which is known to encode the specific heat (or entropy) of charge carriers. In a dual representation, electric current in the transverse direction plays the role of a ``voltage" applying a {\it longitudinal} (Magnus) force on the vortices, which balances the force imposed by the temperature gradient.The latter is proportional to the entropy of the vortices. Hence, $\alpha_{xy}$ vanishes in the case where vortices carry no entropy (see also Ref. \cite{Jnernst}).

We finally note that the remarkably simple relation to the entropy per carrier $\alpha_{xy}\sim -(s/B)$, derived for clean (Galilean invariant) systems \cite{CHR,BO}, does not hold here.
Indeed, this relation can be recovered, e.g., employing the Boltzmann equation for an ordinary conductor in the limit $\omega_c\tau\rightarrow\infty$ (with $\omega_c$ the cyclotron frequency and $\tau$ a mean free time \cite{BoltzmannNotes}). In contrast, for $\omega_c\tau\ll 1$ the same calculation yields $\alpha_{xy}\sim B\tau^2$. In our case, the latter limit is appropriate: while translational invariance holds in the $x$-direction, charge conductance along the $y$-direction is governed by weak tunneling between two discrete points, $\sim J$. We hence expect $\alpha_{xy}\sim BJ^2$, in accord with Eq. (\ref{alpha_xy_general}).

As a concluding remark, we note that a qualitatively similar behavior of the Nernst effect is expected to hold in 2D SC films, or an infinite stack of such ladders (which is essentially equivalent to an anisotropic ultrathin SC film). Possibly, it can also explain some properties of the existing data: see, for example, Fig. 3 in the paper by Pourret {\it et al.} \cite{Pourret}, where the Nernst signal measured in NbSi films exhibits an increase (and sharpening) of the peak for $T<T_c$. It is therefore suggestive that more elaborate Josephson arrays models can serve as a useful arena for studying transverse thermoelectric effects in disorder SC films.

\acknowledgements

We thank A. Auerbach, D. Arovas, K. Behnia, A. Frydman, A. Goldman, K. Michaeli and D. Podolsky for useful discussions. E. S. is grateful to the hospitality of the Aspen Center for Physics (NSF Grant No. 1066293). This work was supported by the US-Israel Binational Science Foundation (BSF) grant 2008256, and the Israel Science Foundation
(ISF) grant 599/10.


\begin{thebibliography}{99}

\bibitem{SCfluc}
For a review see, e.g., A. I. Larkin and A. A. Varlamov in {\it The Physics of Superconductors}, Vol. I, eds. K.-H.Bennemann and J. B. Ketterson (Springer, Berlin 2003).

\bibitem{SITrev} For a review and extensive references, see
%A. F. Hebard, in
%\textit{Strongly Correlated Electronic Materials} (The Los Alamos Symposium
%1993), Eds. K. S. Bedell, Z. Wang, D. E. Meltzer, A. V. Balatsky and E.
%Abrahams, Addison Wesley (1994), p. 251; G. T. Zimanyi, \textit{ibid} p. 285;
Y. Liu and A. M. Goldman, Mod. Phys. Lett. B \textbf{8}, 277 (1994); S.
L. Sondhi, S. M. Girvin, J. P. Carini and D. Shahar, Rev. Mod. Phys. \textbf{
69}, 315 (1997); A. M. Goldman and N. Markovic, Physics Today \textbf{51}, 39
(1998); N. Nagaosa, \textit{Quantum Field Theory in Condensed Matter Physics}, Sec. 5.3 (Springer, 1999).

\bibitem{SIT1D}
K. Yu. Arutyunov, D. S. Golubev and A. D. Zaikin, Physics Reports {\bf 464}, 1
(2008), and refs. therein.

\bibitem{ong1}
Z. A. Xu, N. P. Ong, Y. Wang, T. Kakeshita and S. Uchida, Nature {\bf 406}, 486 (2000); Y. Wang, Z. A. Xu, T. Kakeshita, S. Uchida, S. Ono, Y. Ando and N. P. Ong, Phys. Rev. B {\bf 64}, 224519 (2001).

\bibitem{ong2}
Y. Wang, L. Li and N. P. Ong, Phys. Rev. B {\bf 73}, 024510 (2006).

\bibitem{behnia}
P. Spathis, H. Aubin, A. Pourret and K. Behnia, Europhys. Lett. {\bf 83}, 57005 (2008).

\bibitem{Pourret}
A. Pourret, P. Spathis, H. Aubin and K. Behnia, New J. Phys. {\bf 11}, 055071 (2009).

\bibitem{UD}
S. Ullah and A. T. Dorsey, Phys. Rev. Lett. {\bf 65}, 2066 (1990); S. Ullah and A. T. Dorsey, Phys. Rev. B {\bf 44}, 262 (1991).

\bibitem{USH}
I. Ussishkin, S. L. Sondhi and D. A. Huse, Phys. Rev. Lett. {\bf 89}, 287001 (2002).

\bibitem{SSVG}
M. N. Serbyn, M. A. Skvortsov, A. A. Varlamov and V. M. Galitski, Phys. Rev. Lett. {\bf 102}, 067001 (2009); A. Sergeev, M. Reizer, and V. Mitin, Phys. Rev. Lett. {\bf 106}, 139701 (2011); M. N. Serbyn, M. A. Skvortsov, A. A. Varlamov and V. M. Galitski, Phys. Rev. Lett. {\bf 106}, 139702 (2011).

\bibitem{MF}
K. Michaeli and A. M. Finkelstein, Europhys. Lett. {\bf 86}, 27007 (2009).

\bibitem{KT}
J. M. Kosterlitz and D. J. Thouless, J. Phys. C {\bf 6}, 1181 (1973).

\bibitem{PRV}
D. Podolsky, S. Raghu and A. Vishwanath, Phys. Rev. Lett. {\bf 99}, 117004 (2007).

\bibitem{BGS}
M. J. Bhaseen, A. G. Green and S. L. Sondhi, Phys. Rev. Lett. {\bf 98}, 166801 (2007); Phys. Rev. B {\bf 79}, 094502 (2009).

\bibitem{CHR}
N. R. Cooper, B. I. Halperin and I. M. Ruzin, Phys. Rev. B {\bf 55}, 2344 (1997).

\bibitem{BO}
D. L. Bergman and V. Oganesyan, Phys. Rev. Lett. {\bf 104}, 066601 (2010).

\bibitem{SRM}
A. Sergeev, M. Reizer and V. Mitin, Europhys. Lett. {\bf 92}, 27003 (2010).

\bibitem{glazman}
Devices of this type have been already fabricated and utilized to study magneto-resistance due to quantum dynamics of vortices; see, e.g., C. Bruder, L.I. Glazman, A.I. Larkin, J.E. Mooij and A. van Oudenaarden, Phys. Rev. B \textbf{59}, 1383 (1999).

\bibitem{Jnernst}
The Nernst effect in a classical version of such device - a long Josephson junction between two bulk SC - has been measured and studied in detail in earlier literature; see, e.g., G. Yu. Logvenov, V. A. Larkin and V. V. Ryazanov, Phys. Rev. B {\bf 48}, 16853(R) (1993); V. M. Krasnov, V. A. Oboznov and N. F. Pedersen, Phys. Rev. B {\bf 55}, 14486 (1997).

\bibitem{3rd_law}
We note that this behavior is confined to an intermediate regime of $T$, where our approximations hold (see Sec. II), and possibly changes in the limit $T\rightarrow 0$.

\bibitem{Giamarchi} See, e.g., T. Giamarchi, \textit{Quantum Physics in One Dimension},
(Oxford, New York, 2004).

\bibitem{OG} E. Orignac and T. Giamarchi, Phys. Rev. B \textbf{64}, 144515
(2001).

\bibitem{AS} Y. Atzmon and E. Shimshoni, Phys. Rev. B \textbf{83}, 220518(R)
(2011); Phys. Rev. B \textbf{85}, 134523 (2012).

\bibitem{GS}
T. Giamarchi and H. J. Schulz, Phys. Rev. B {\bf 37}, 325 (1988).

\bibitem{ESA}
U. Eckern, G. Sch{\"o}n and V. Ambegaokar, Phys. Rev. B {\bf 30}, 6419 (1984).

\bibitem{SGF}
E. Shimshoni, Y. Gefen and S. Fishman, Phys. Rev. B {\bf 40}, 2158 (1989).

\bibitem{RA}
A. Rosch and N. Andrei, Phys. Rev. Lett. {\bf 85}, 1092 (2000).

\bibitem{GRbook}
I. S. Gradshteyn and I. M. Ryzhik, \textit{Tables of Integrals, Series and Products} (Academic Press, 1980).

\bibitem{zaikin}
A. D. Zaikin, D. S. Golubev, A. van Otterlo and G. T. Zimanyi,
Phys. Rev. Lett.
{\bf 78}, 1552 (1997).

\bibitem{BoltzmannNotes}
Yeshayahu Atzmon, \textit{Superconductivity in low-dimensional systems}, Ph.D. thesis (Bar-Ilan University, 2012); see Appendix C.

\end{thebibliography}
\end{document}